\documentclass{article}
\usepackage{spconf}
\usepackage{cite}
\usepackage{amsmath,amssymb,amsfonts}
\usepackage{algorithmic}
\usepackage{academicons}
\usepackage{graphicx}
\usepackage{textcomp}
\usepackage{xcolor}
\usepackage{todonotes}
\usepackage{caption}
\usepackage{subcaption}
\usepackage[hyphens]{url}
\usepackage{hyperref}
\usepackage[hyphenbreaks]{breakurl}
\usepackage{balance}
\usepackage{todonotes}
\graphicspath{{./images/}}
\title{L3DAS22 CHALLENGE: LEARNING 3D AUDIO SOURCES \\IN A REAL OFFICE ENVIRONMENT}
\makeatletter
\def\@name{

\emph{Eric~Guizzo$^{\dagger}$},
\emph{Christian~Marinoni$^{\dagger}$}, \emph{Marco~Pennese$^{\dagger}$},
\emph{Xinlei~Ren$^{\ast}$}, 
\emph{Xiguang~Zheng$^{\ast}$}, \\
\emph{Chen~Zhang$^{\ast}$},
\emph{Bruno~Masiero$^{\star\star}$}, 
\emph{Aurelio~Uncini$^{\dagger}$},
\emph{Danilo~Comminiello$^{\dagger}$}

\thanks{Corresponding author's email: \href{mailto:danilo.comminiello@uniroma1.it}{danilo.comminiello@uniroma1.it}. This work has been performed while the first author was a PhD Visiting Student with Sapienza University of Rome. This work has been partly supported by ``Progetti di Ricerca Grandi'' of Sapienza University of Rome under grant number RG11916B88E1942F and the S\~{a}o Paulo Research Foundation (FAPESP) under grant number 2017/08120-6.}\vspace{1em}}
\makeatother
\address{$^{\dagger}$DIET Dept., Sapienza University of Rome, 
Italy \\ 
$^{\ast}$Kuaishou Technology Co., 
Beijing, China \\
$^{\star\star}$DECOM Dept., University of Campinas, 
Brazil}
\begin{document}
\ninept
%
%
\maketitle
%
%
%
\begin{abstract}
The L3DAS22 Challenge 
is aimed at encouraging the development of machine learning strategies for 3D speech enhancement and 3D sound localization and detection in office-like environments. This challenge improves and extends the tasks of the L3DAS21 edition\footnote{\url{www.l3das.com/mlsp2021}}.
We generated a new dataset, which maintains the same general characteristics of L3DAS21 datasets, but with an extended number of data points and adding constrains that improve the baseline model's efficiency and overcome the major difficulties encountered by the participants of the previous challenge.
We updated the baseline model of Task 1, using the architecture that ranked first in the previous challenge edition. 
We wrote a new supporting API, improving its clarity and ease-of-use.
In the end, we present and discuss the results submitted by all participants. L3DAS22 Challenge website: {\url{www.l3das.com/icassp2022}}.


%
\end{abstract}
%
%
%
%
%
%
\section{Introduction}
\label{sec:intro}
Machine learning applications of 3D audio are gaining increasing interest in recent years. 
Tasks like sound source localization, sound source separation, speech enhancement and acoustic echo cancellation, among others, potentially benefit from tridimensional representations of sound field, as they carry additional spatial information \cite{abesser2020review, DBLP:conf/ijcnn/AdavannePV18}.
Consequently, in these tasks 3D audio formats (in particular Ambisonics) usually provide performance improvements compared to single/dual-channel formats \cite{DBLP:conf/ijcnn/AdavannePV18, DBLP:conf/icassp/AdavannePV17}. 
Based on this motivation, the L3DAS (Learning 3D Audio Sources) Team has proposed the L3DAS22 Challenge involving two tasks, 3D Speech Enhancement (SE) and 3D Sound Event Localization and Detection (SELD), both relying on multiple-source and multiple-perspective (MSMP) Ambisonics recordings. 

3D SE aims at removing unwanted information from spurious spatial vocal recordings and further enhancing the speech intelligibility and clarity. 
A widespread strategy to perform SE is to use deep neural networks to estimate a mask in the Time-Frequency domain that tries to remove unwanted noise components from the signal mixture \cite{DBLP:journals/taslp/WangC18a}.
Neural beamforming techniques, such as the Filter and Sum Networks (FaSNet) \cite{DBLP:conf/asru/LuoHMCL19}, provide state-of-the art results for Ambisonics-based SE and are usually suitable for low-latency scenarios. 
Also U-Net-based approaches provide competitive results both for monaural  
\cite{DBLP:journals/eswa/GuimaraesNS20}
and multichannel SE tasks 
\cite{xinleiref}, at the expense of higher computational power demand.
Other techniques to perform SE include recurrent neural networks (RNNs) \cite{DBLP:conf/icassp/HuangKHS14}, graph-based spectral subtraction \cite{DBLP:journals/speech/YanYWG20}, discriminative learning \cite{DBLP:conf/interspeech/FanLTYW19}, and dilated convolutions
\cite{DBLP:journals/taslp/LuoM19}, among others.

3D SELD, instead, aims at obtaining exhaustive spatiotemporal descriptions of 3D acoustic scenes, predicting which sound categories are present in the scene, and when and where each sound instance is active. 
SELD can be considered as a combination of the traditional sound event detection and sound source localization tasks, and it was presented for the first time in the DCASE2019 Challenge \cite{DBLP:journals/taslp/PolitisMAHV21}.
Also here, the state-of-the-art methods are based on deep learning strategies \cite{DBLP:journals/corr/abs-2006-01919}. SELDnet \cite{DBLP:journals/jstsp/AdavannePNV19} adopted a convolutional-recurrent design with two distinct branches for localization and detection and it was used as a baseline model in SELD tasks of the DCASE challenges.
An improved SELDnet model was then introduced by \cite{DBLP:conf/eusipco/GuirguisSGAY20}, including temporal convolutions.
Other novel solutions for this task include ensemble models \cite{chytas2019hierarchical}, multi-stage training \cite{DBLP:journals/corr/abs-1905-00268} and bespoke augmentation strategies
\cite{DBLP:journals/corr/abs-1910-04388}.

In this second edition, we improved many aspects of the L3DAS21 challenge \cite{guizzo2021l3das21}. 
First of all, we generated a new dataset (L3DAS22 dataset\footnote{The L3DAS22 Dataset is freely available on Kaggle: \url{www.kaggle.com/l3dasteam/l3das22}}) with an augmented number of datapoints, increasing the total dataset duration from 65 to more than 94 hours. 
Moreover, we analyzed the major difficulties encountered by the participants of the previous edition and we modified the dataset synthesis pipeline in order to promote less resource-demanding trainings and facilitate both tasks.
In addition, we propose an updated baseline for task 1, using the model architecture that ranked first in the previous edition, which provides an improved baseline metric of 0.81 (previously 0.62).
Finally, we rewrote the supporting API, fixing existing bugs and making clearer and faster the preprocessing and baseline training/evaluation stages.

%

The rest of the paper is organized as follows: Section \ref{sec:data} exposes the details of the L3DAS22 datasets for both the tasks of 3D SE and 3D SELD. Section \ref{sec:tasks} describes the challenge tasks, while in Section \ref{sec:reseval} we illustrate the details of the baseline models. Section \ref{sec:rules} contains information on the challenge conduct and Section \ref{sec:results} discusses the submission results. Finally, Section \ref{sec:conclusion} draws the conclusions of this paper.
%
%
%
%
%
\section{A 3D AUDIO DATASET FROM A REAL REVERBERANT OFFICE ENVIRONMENT}
\label{sec:data}

\subsection{3D Impulse Response Recording and Data Collection}
\label{sec:collection}
\begin{figure}[t]
 \centering
 \includegraphics[width=0.48\textwidth,keepaspectratio]{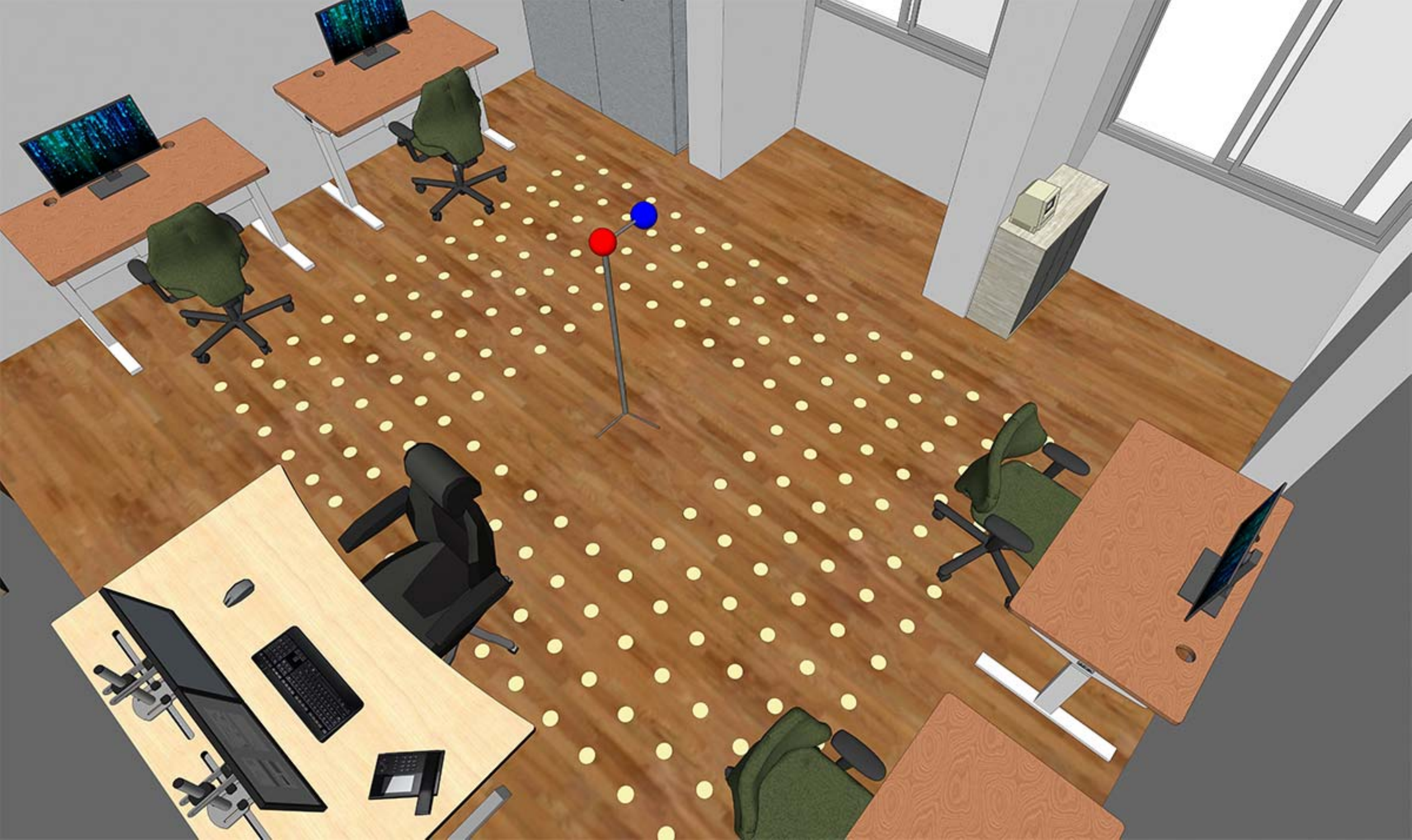}
\caption{3D representation of the office used for the recordings. The red sphere represents microphone A, while the blue one represents microphone B.}
\label{fig:3d-view}
\end{figure}

The L3DAS22 dataset contains approximately 98 hours of MSMP B-format  audio recordings. 
%
We sampled the acoustic field of a real office room with the approximate dimensions of 6 m (length) by 5 m (width) by 3 m (height). 
The room has typical office furniture, a wooden parquet floor and painted concrete walls and ceiling. We used two first-order A-format Soundfield Ambisonics microphones\footnote{Oktava MK-4012}, one placed in the exact center of the room (mic A) and the other 20 cm distant towards the width dimension (mic B), as shown in Figure \ref{fig:3d-view}. We positioned both microphones at the same height of 1.3 m, which is the average ear height of a seated person. Moreover, their capsules have the same orientation.

\begin{figure}[t]
 \centering
 

    \begin{subfigure}{0.5\linewidth}
    \includegraphics[width=\linewidth]{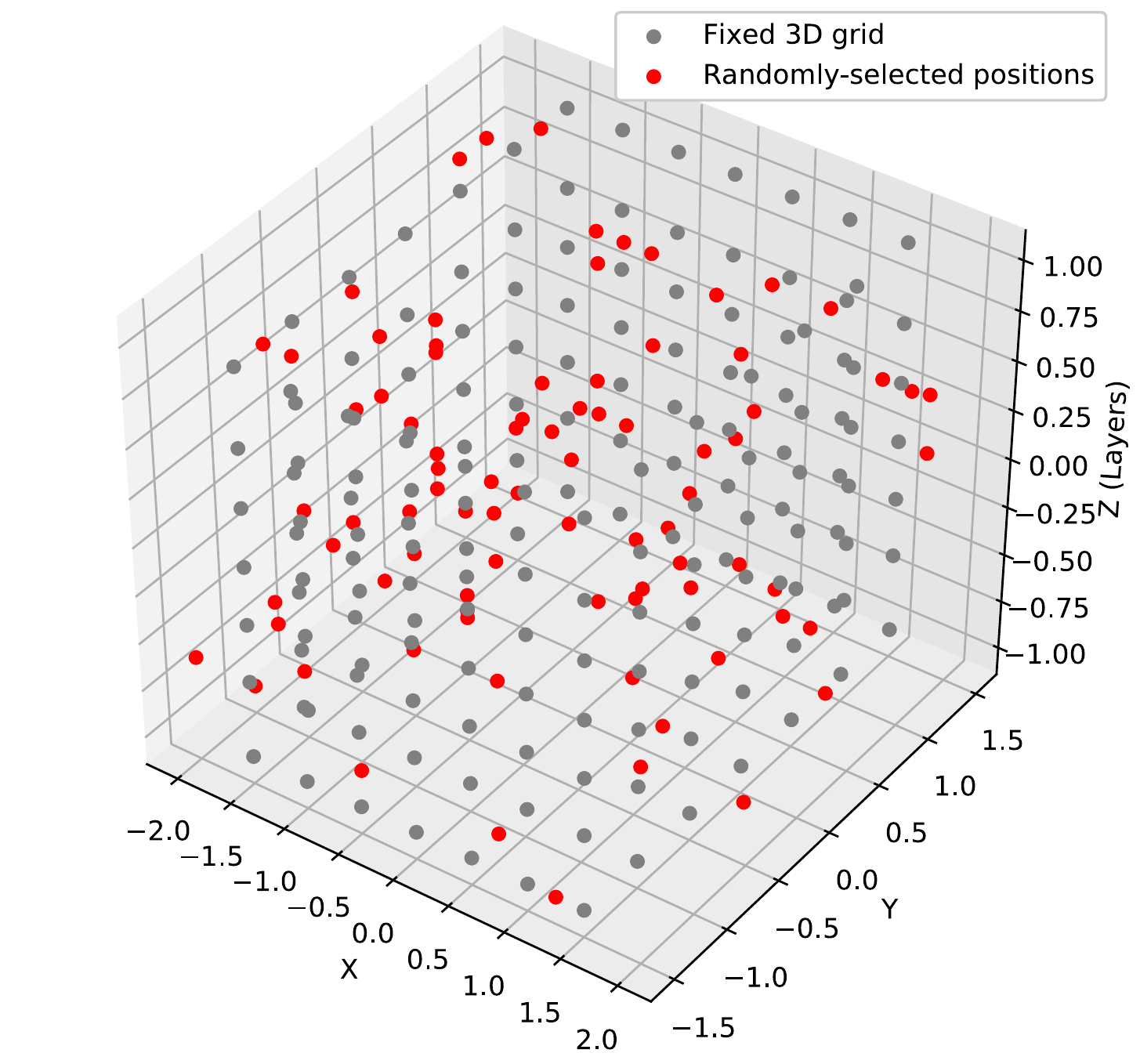}

    \caption{3D speaker positions}
    \label{fig:3dpositions}
    \end{subfigure}
    \begin{subfigure}{0.48\linewidth}
    \includegraphics[width=\linewidth]{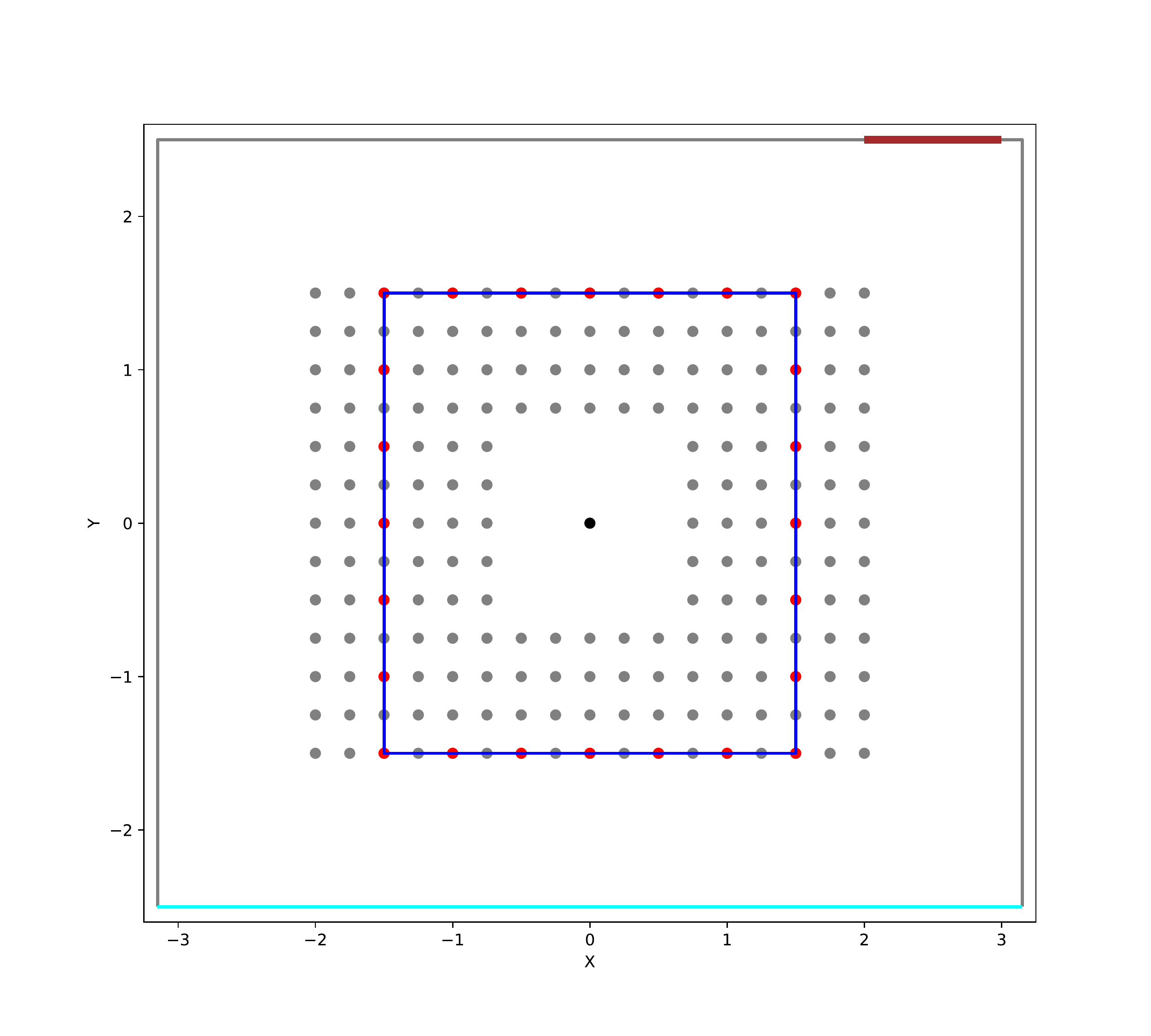}

    \caption{Grid projection}
    \label{fig:gridpositions}
    \end{subfigure}\hfill
    
    \caption{(a) Tridimensional distribution of the speaker positions. In gray the fixed 3D grid, in red the distribution of the randomly-selected positions.
    (b) Projection from above of the microphones position (center dot) and the speaker positions of the fixed 3D grid (red dots connected by the blue line). }
    \label{fig:totalpositions}
\label{fig:positions}
\end{figure}

We reproduced an analytic signal using a speaker\footnote{Event PS6} placed in 252 fixed spatial positions chosen according to two criteria: a fixed 3D grid (168 positions) and a 3D uniform random distribution (84 positions).
Figure \ref{fig:gridpositions} shows a 2D projection of the grid from above. Given the first criterion, we placed the speaker in a 3D grid with a 50 cm step in the length-width dimensions, as represented in Fig.~\ref{fig:gridpositions} with gray dots.
In the height dimension, we considered 7 position layers at 0.3 m, 0.7 m, 1 m, 1.3 m, 1.6 m, 1.9 m and 2.3 m from the floor, as shown in Figure \ref{fig:3dpositions}. 
On the other hand, the random positions are uniformly sampled among those available in a virtual 3D grid having a 25 cm step and are depicted in red in Fig.~\ref{fig:3dpositions}.
For all measurements, we directed the speaker's tweeter towards mic A by changing the incline of its support.

The analytic signal is a 24-bit exponential sinusoidal sweep that glides from 50 Hz to 16000 Hz in 20 seconds, reproduced at 90 dB SPL on average.
The IR estimation is then obtained by performing a circular convolution between the recorded sound and the time-inverted analytic signal, as introduced by \cite{farina2000simultaneous}.
We finally converted the A-format signals into standard B-format IRs\footnote{\url{http://pcfarina.eng.unipr.it/Public/B-format/A2B-conversion/A2B.htm}}.

We considered the collected Ambisonics impulse responses and some existing clean monophonic datasets, and we applied a convolution operation to virtually place that sound source in the spatial position occupied by the speaker, as perceived from the 2 microphones. The result is a set of synthetic tridimensional sound sources obtained by convolving the original sounds with our IRs.
We aimed at creating plausible and diverse 3D scenarios to reflect office-like situations, in which disparate types of sound sources and background noises coexist in the same 3D reverberant environment.

For this purpose, we used the Librispeech \cite{DBLP:conf/icassp/PanayotovCPK15} and FSD50K \cite{DBLP:journals/corr/abs-2010-00475} 
datasets.
More precisely, we selected a total of 1440 noise sound files from FSD50K equally distributed between 14 transient noise classes:\textit{ computer keyboard, drawer open/close, cupboard open/close, finger snapping, keys jangling, knock, laughter, scissors, telephone, writing, chink and clink, printer, female speech, male speech}, and 4 continuous noise classes: \textit{alarm, crackle, mechanical fan, microwave oven}.
Furthermore, we extracted clean speech signals (without background noise) from Librispeech, taking only sound files up to 12 seconds.

Our dataset is partitioned into two sections, each of which is dedicated to a different challenge task. As predictor data for both tasks, we supply normalized raw waveforms of all Ambisonics channels (8 signals in total), whereas the target data differs significantly.
Also, we developed a variety of acoustic scenarios that were tailored to each task.
%


\subsection{L3DAS22 Dataset for Task 1: 3D SE}
\label{sec:dataset_se}
For the task 1, related to 3D SE, we synthesized more than 40000 virtual 3D audio environments, for a total length of approximately 90 hours. In each data point a speech signal with a duration up to 12 seconds is always present, mixed with various types of background noise. 
We extracted all the speech sounds from the clean subset of Librispeech (approximately 53\% male and 47\% female speech).
We added up to 3 simultaneous non-speech background noises of the above-mentioned categories, extracting them from FSD50K.
With a 25\% chance, one of the background noises is a continuous noise.
The signal-to-noise ratio ranges from 6 to 16 dBFS (referring to the signals' RMS amplitude), where the voice is always the prominent signal. We randomly placed all sound sources in the 3D environment, paying attention to obtain a uniform distribution of locations within this dataset section.

The predictors data for this task are released as 8-channel 16 kHz 16 bit wav files, consisting of 2 sets of first-order Ambisonics recordings.
The channels order follows the Ambisonics Channel Number (ACN) system of the AmbiX format \footnote{\url{http://pcfarina.eng.unipr.it/aurora/B-Format_to_UHJ.htm}}, thus having [WA, YA, ZA, XA, WB, YB, ZB, XB], where the letters A, B, refer to the used microphone and W, Y, Z, X, refers to the B-format Ambisonics channels.
The target data provided contains the clean monophonic recordings of the only speech signals (16 kHz 16 bit mono wav files), as well as the words uttered in each data point (in a txt file).

For this task, we have also provided an informative csv file for each subset, where we annotated the coordinates and spatial distance of the IR convolved with the target voice signals for each datapoint.
This may be useful to estimate the delay caused by the virtual time-of-flight of the target voice signal and to perform a sample-level alignment of the input and ground truth signals.

\subsection{L3DAS22 Dataset for Task 2: 3D SELD}
\label{sec:dataset_seld}\

For the task 2, related to 3D SELD, we synthesized 900 30-seconds-long data points, reaching a total length of 7.5 hours of audio. 
Each data point contains a simulated 3D office audio environment in which up to 3 simultaneous acoustic events may be active at the same time.
Moreover, when multiple sounds are active at the same time, the probability of the sounds to belong to the same class is artificially increased. As a result, in the case of a maximum of 3 overlaps, two simultaneous sounds may belong to the same class with an approximate probability of 15\% when the overlapped events are 2 or 22\% when 3 sounds are overlapped.
Although, when this happens, we forced the simultaneous sounds of the same class to be virtually positioned at least 1 meter distant from each other. 

The tracks with 1, 2 and 3 overlaps contain an average of 7, 13 and 20 acoustic events, respectively with a standard deviation of 2, 3 and 4.
The sound events belong to the aforementioned 14 transient noise classes and are therefore 1120 in total. 
As opposed to the SE dataset, here the data points are not forced to contain speech signals, although they may contain voice sounds. The volume difference between the different sounds ranges from 0 to 20 dBFS (referring to the signal's RMS amplitude).
Also here, we randomly place all sound sources in the 3D environment, paying attention to obtain a uniform distribution of locations.

The predictors data for task 2 have the same form as for the task 1, except for the sampling frequency, which here is 32 kHz.
As target data, we provide a csv file containing the onset and offset time stamps, the typology class and the spatial coordinates of each individual sound event present in a data point.

\subsection{Dataset Splits}
\label{sec:dataset_splits}\
We split both dataset sections into a training set (approximately 80 hours for SE and 5 hours for SELD) and a test set (approximately 7 hours for SE and 2.5 hours for SELD), paying attention to create similar distributions. 
The train set of the SE section is divided in two partitions: train360 and train100, and contain speech samples extracted from the correspondent partitions of Librispeech (only the samples up to 12 seconds). 
All sets of the SELD section are divided in: OV1, OV2, OV3. 
These partitions refer to the maximum amount of possible overlapping sounds, which are 1, 2 or 3, respectively. 

The test set of both dataset sections is further split into two equally-long subsets that present a similar distribution: one development and one blind test set. 
The first one is part of the initial release of the dataset, and it is aimed, as usual, at the model's hyperparameters fine-tuning. 
The latter, instead, is aimed at the submissions' evaluation and was initially released with the only predictors data, without target labels/signals.

\section{CHALLENGE TASKS}
\label{sec:tasks}

We propose 2 different tasks, both based on our L3DAS22 dataset: \textit{3D Speech Enhancement in Office Reverberant Environment} and \textit{3D Sound Event Localization and Detection in Office Reverberant Environment}.
Each one is divided in 2 sub-tasks: one-mic and dual-mic recordings, respectively relying on the sounds acquired by one or both Ambisonics microphones, as described in Section \ref{sec:data}.

In this context, the information predicted for one task may be beneficial for the other one.
For instance, the sound localization parameters may be re-used to improve the performance of 3D speech enhancement networks, as in \cite{DBLP:conf/eusipco/ChazanHHGG19, DBLP:journals/corr/abs-2010-11566}. 
Therefore, participants are encouraged to develop a strategy to bootstrap the resources and exploit the output of one model to enhance the performance of the other one (although this is not mandatory).

%
\subsection{Description and Goals of Task 1: 3D SE}
\label{subs:speechenh}
The objective of this task is the separation and enhancement of speech signals immersed in a noisy 3D environment, basing on the SE section of the L3DAS22 dataset.
Here the models are expected to extract the monophonic voice signal from the 3D mixture that contains various background noises.  
The evaluation metric for this task is a combination of the short-time objective intelligibility (STOI), which estimates the intelligibility of the output speech signal, and word error rate (WER), computed to assess the effects of the enhancement for speech recognition purposes.
We use a Wav2Vec \cite{DBLP:conf/nips/BaevskiZMA20} architecture pre-trained on Librispeech 960h\footnote{\url{https://huggingface.co/facebook/wav2vec2-base-960h}} to compute the WER.
The final metric for this task is a combination of these two measures given by \((STOI+(1-WER))/2\).
This metric lies therefore in the 0-1 range and higher values are better.

\subsection{Description and Goals of Task 2: 3D SELD}
\label{subs:seld}
The aim of this task is to detect the temporal activity, spatial position and typology of a known set of sound events immersed in a synthetic 3D acoustic environment.
This task is performed on the SELD section of the L3DAS22 dataset.
Here the models are expected to predict a list of the active sound events and their respective location at regular intervals of 100 milliseconds.

We use a joint metric for localization and detection: location-sensitive detection error, as defined in \cite{DBLP:conf/waspaa/MesarosAPHV19}. 
This metric is computed on each time frame and consists of measuring the Cartesian distance between the predicted and true events with the same label, and counting a true positive only when its label is correct and its location is within a threshold from its reference location. 
After this operation, we compute the regular F score. 
Since the scenario is particularly complex and challenging, we fixed the spatial error threshold to 2 meters for this task.

\section{BASELINE METHODS}
\label{sec:reseval}
As baseline methods, we propose state-of-the-art architectures, specifically adapted for each task. For both tasks, we used only signals coming from one Ambisonics microphone (mic A), leaving room for experimentation with the dual-mic configuration.



For task 1 (SE), we use a beamforming U-Net architecture 
\cite{xinleiref}, 
which  provided the best metrics for the L3DAS21 Challenge on the SE task. 
This network uses a convolutional U-Net to
estimate B-format beamforming filters and contains three
main modules: encoder for extracting high-level features, decoder for reconstructing the size of input features from the output of the encoder, and skip connections for
concatenating each layer in the encoder with its corresponding layer in the decoder.
The enhancement process is performed as that of the traditional signal beamforming. 
We multiply the complex spectrogram of B-format noisy signal with the filters estimated by U-Net through element-wise multiplication, and
then sum the result over the channel axis to estimate a single-channel enhanced complex spectrogram. 
In the end the ISTFT is performed to obtain the enhanced time-domain signal.
With this model we obtained a baseline test metric for task 1 of 0.83, with a word error rate of 0.21 and a STOI of 0.88.

For task 2, instead, we developed a variant of the SELDnet architecture \cite{DBLP:journals/jstsp/AdavannePNV19}.
We ported to the PyTorch language the original Keras implementation\footnote{\url{https://github.com/sharathadavanne/seld-net}} and we modified its structure in order to make it compatible with the L3DAS22 dataset. 
The objective of this network is to output a continuous estimation (within a fixed temporal grid) of the sounds present in the environment and their respective location.
The original SELDNet architecture is conceived for processing sound spectrograms (including both magnitudes and phase information) and uses a convolutional-recurrent feature extractor based on 3 convolution layers followed by a bidirectional GRU layer. 
In the end, the network is split in two separate branches that predict the detection (which classes are active) and location (where the sounds are) information for each target time step.
We augmented the capacity of the network by increasing the number of channels and layers, while maintaining the original data flow.
Moreover, we discard the phase information and we perform max-pooling on both the time and the frequency dimensions, as opposed to the original implementation, where only frequency-wise max-pooling is performed.
In addition, we added the ability to detect multiple sound sources of the same class that may be active at the same time (3 at maximum in our case). 
To obtain this behavior we tripled the size of the network's output matrix, in order to predict separate location and detection information for all possible simultaneous sounds of the same class.
This network obtains a baseline test F-score of 0.34, with a precision of 0.42 and a recall of 0.29.

For further implementation details on our baseline models, please refer to the L3DAS official GitHub repository\footnote{\url{https://github.com/l3das/L3DAS22}}.

\section{RULES AND CONDUCT OF THE CHALLENGE}
\label{sec:rules}
The L3DAS22 Challenge lasted 8 weeks, from the release to the submission date. All the participants were allowed to submit results for at least one of the two challenge tasks. Each individual participant was not allowed to join more teams, thus having the possibility to submit only one set of results.

No restrictions were placed on the methods to be used for the two tasks. Teams had the possibility to choose their best results among those obtained in the 1-mic and 2-mic configurations. It was also allowed to augment the L3DAS22 dataset and/or to integrate additional data with pretrained models.

Challenge winners have been selected according to the best performance for each task, separately.

\section{CHALLENGE RESULTS}
\label{sec:results}

\begin{figure}[t]
 \centering
 \begin{subfigure}{\linewidth}
    \includegraphics[width=\linewidth]{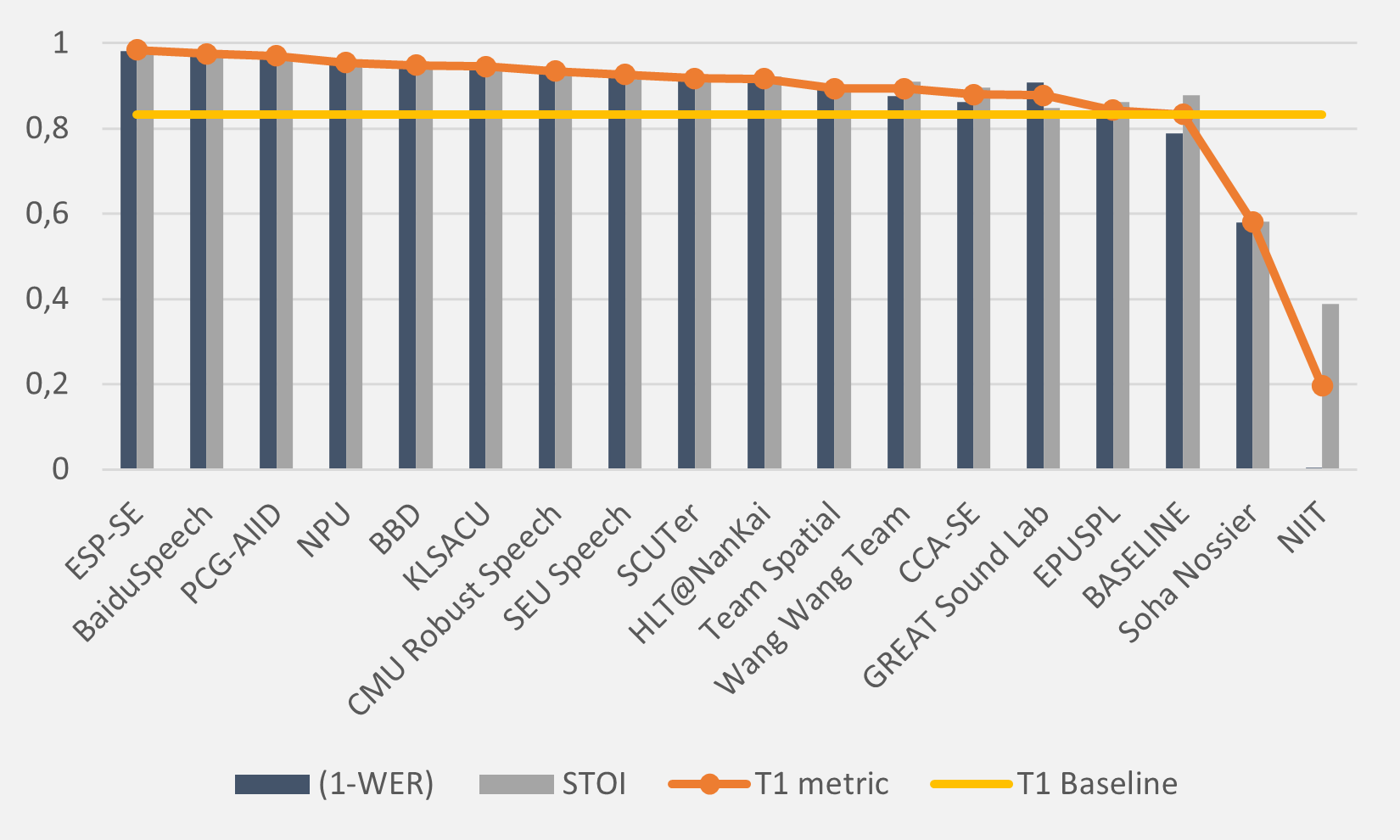}
    \caption{}
    \label{fig:resultsT1}
    \end{subfigure}
    \begin{subfigure}{\linewidth}
    \includegraphics[width=\linewidth]{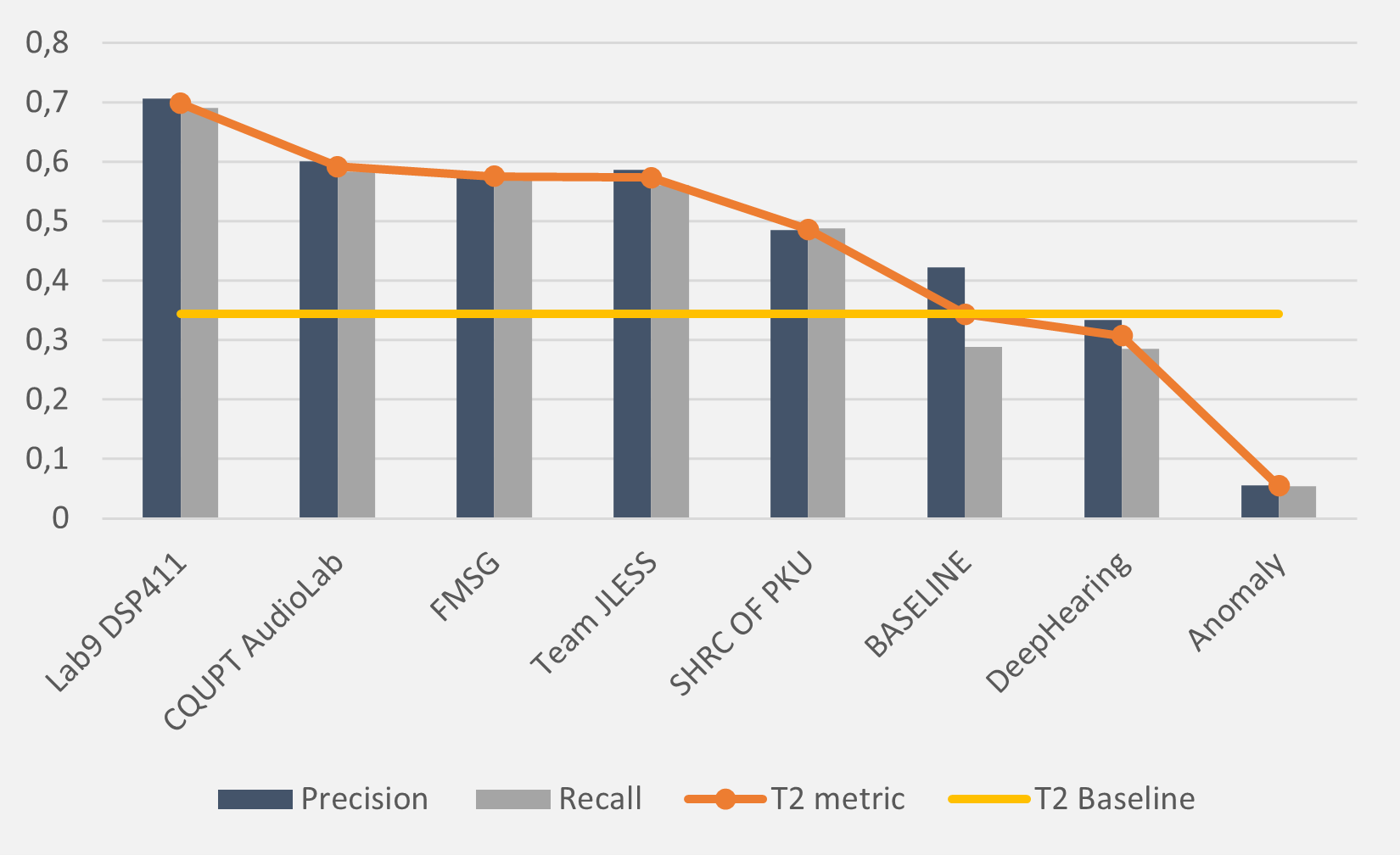}
    \caption{}
    \label{fig:resultsT2}
    \end{subfigure}
    \caption{L3DAS22 Challenge results for (a) Task 1: 3D SE, and (b) Task 2: 3D SELD.}
    \label{fig:results}
\end{figure}

The L3DAS22 Challenge has received 46 registrations and 24 result submissions: 17 teams submitted results for the task 1 and 7 teams for task 2. A graphical illustration of the results has been reported in Fig. \ref{fig:results}, where it can be seen that several teams have improved the baseline results.

In particular, the winner team for Task 1, ESP-SE, has obtained a T1 metric score of 0.984, with a WER of 0.019 and a STOI of 0.987. On the other hand, the winner team for Task 2, Lab9\_DSP411, has obtained a T2 metric score of 0.699, with a precision of 0.706 and a recall of 0.691. Further information about challenge results and awards can be found on the L3DAS22 Challenge website\footnote{\url{www.l3das.com/icassp2022/results}}.

\section{CONCLUSION}
\label{sec:conclusion}
This paper has introduced a Signal Processing Grand Challenge, named L3DAS22 Challenge: Machine Learning for Signal Processing. Alongside the challenge we presented a new dataset on 3D audio recorded in a real reverberant office environment and two different tasks on 3D SE and 3D SELD. 
Future works by the L3DAS Team will involve more challenging 3D acoustic scenarios, different microphone configurations and also new tasks.

%
%
\clearpage
\balance
\bibliographystyle{IEEEbib-abbrev}
\ninept
\bibliography{ICASSP22refs}
%
%
\end{document}